\documentclass{article}
\usepackage{nicefrac}
\usepackage{amsfonts}
\usepackage{amssymb}
\usepackage{amsmath,mathrsfs,dsfont}
\usepackage{graphicx}
\usepackage{appendix}
\usepackage{cite}
\newcommand{\D}{D\!\!\!\!/}
\newcommand{\Tr}{\mathrm{Tr}}

\begin{document}

\title{On spectral geometry approach to Ho\v{r}ava-Lifshitz gravity: Spectral dimension.}
\author{A. Pinzul\thanks{apinzul@unb.br} \\
\\
\emph{Universidade de Bras\'{\i}lia}\\
\emph{Instituto de F\'{\i}sica}\\
\emph{70910-900, Bras\'{\i}lia, DF, Brasil}\\
\emph{and}\\
\emph{{International Center of Condensed Matter Physics} }\\
\emph{C.P. 04667, Brasilia, DF, Brazil} \\
}
\date{}
\maketitle

\begin{abstract}
We initiate the study of Ho\v{r}ava-Lifshitz models of gravity in the framework of spectral geometry. As the first step, we calculate the dimension of space-time. It is shown, that for the natural choice of a Dirac operator (or rather corresponding generalized Laplacian), which respects both the foliation structure and anisotropic scaling, the result of Ho\v{r}ava on a spectral dimension is reproduced for an arbitrary, non-flat space-time. The advantage and further applications of our approach are discussed.
\end{abstract}

\newpage

\section{Introduction}

In \cite{Horava:2009uw}, Ho\v{r}ava used the idea of the anisotropic scaling of the Lifshitz model \cite{Lifshitz} to construct a possible UV completion of Einstein gravity. Ho\v{r}ava's approach has attracted considerable attention from the scientific community. And though the original model suffers from several problems, both theoretical and phenomenological, see, e.g., \cite{Sotiriou:2010wn} for the discussion as well as a nice introduction to the Ho\v{r}ava-Lifshitz gravity, the possibility of having a theory of quantum gravity remaining in the framework of conventional quantum field theory keeps interest in this type of models at a very high level. Several extensions of the original proposal have been suggested to (partially) cure its problems \cite{Blas:2010hb}. One of the main downsides of these extensions, from our point of view, is a large number of marginal and relevant terms. In some models it goes up to a hundred and even could involve an arbitrary function of the lapse $N$ (for details see, e.g., \cite{Blas:2010hb}). To gain some control over this, Ho\v{r}ava suggested (even before the extensions have been introduced) that a detailed balance condition should be imposed \cite{Horava:2009uw}, which could drastically limit the number of terms. This turned out to be too restrictive (and not very well motivated), because models with detailed balance have no chance of passing experimental tests, see \cite{Mukohyama:2009zs} on the phenomenology of the HL gravity.

In this paper, we suggest a different approach to HL-type theories. We propose to analyze these models from the spectral geometry point of view. From the geometrical perspective, the key difference between Einstein's General Relativity and the HL model is that we now have a \textit{foliated} manifold instead of just a manifold. It is well known that the geometrical information, in the case of the usual differential and Riemannian geometry, can be completely recovered from the knowledge of a standard Dirac operator (plus an algebra of functions and a Hilbert space, but these are needed already on the level of topology) \cite{GraciaBondia:2001tr}. When some additional structure, as foliation, is present the corresponding Dirac operator should respect it. Of course, this is true for the standard choice, $\D = i\gamma^\mu (\partial_\mu + \omega_\mu)$, but while recovering the geometry from it, the information about the foliation will be lost. To correct this, we have to make a choice: we can use Dirac operators compatible with the foliation structure constructed out of the standard Dirac operator for the underlying manifold (see, e.g. \cite{Glazebrook}), or we can employ a more intuitive (physical) approach by choosing our Dirac operator based on some physical motivations (of course, it still has to respect the foliation structure!). We will argue that the second approach seems to be more relevant for the analysis of HL-type theories. In this work, we initiate the spectral geometry analysis of HL models. As the first step, we analyze using the methods of spectral geometry the observation that the effective dimension of space-time in HL gravities changes from $d=4$ to $d=2$ while we go from IR to UV physics \cite{Horava:2009if}.

What advantage can provide the spectral geometry in the analysis of HL models? Taking into account the mathematical complexity of this approach, to be justified, it should really improve our understanding of HL-type gravities in addition to just reproduction of known results. In our opinion, this approach has the potential to do this.

Firstly, using methods of the spectral geometry, we put HL gravities back into the geometrical framework. From this point of view, HL gravity is not that different from General Relativity: both are dynamical theories of some underlying geometry. What is different is this geometry, which is now not a (pseudo)-Riemann manifold but a foliation. The spectral geometry has the most effective tools to study such generalized geometries. For example, in this paper we demonstrate how using techniques of the spectral geometry the result of \cite{Horava:2009if} on UV/IR dimensions can be generalized to arbitrary non-flat space-times.

Secondly, we believe that our approach might help to get a better control over the ambiguities of the action for HL models. The hope is that this could be achieved through the so-called spectral action \cite{Chamseddine:1996zu}. This action depends only on the spectrum of some Dirac operator and for the standard choice of this operator it leads exactly to the Einstein-Hilbert action of General Relativity \cite{Chamseddine:1996zu}. Because the choice of a Dirac operator for a foliated manifold is much more restrictive then the choice of the action,\footnote{Essentially, we have to write all possible Dirac operators up to the third order and respecting the foliation structure. This still has a lot of freedom, but significantly less then in writing an action directly.} we expect that the spectral action will have less free parameters.

Thirdly, the spectral geometry approach to physics (in the form of non-commutative geometry \cite{Connes:1994yd}) has proven very effective in deriving the Standard Model from a purely geometrical point of view \cite{Chamseddine:2008zj}. In the same spirit, our approach should provide a natural way to couple HL gravity to gauge and matter fields.

We elaborate on some of the points above in Section {\bf 5}.

The paper is organized as follows: in Section {\bf 2} we briefly review Ho\v{r}ava's approach to calculating UV/IR dimensions of the model; Sections {\bf 3} and \textbf{4} contain the main result of this letter - the calculation of the spectral dimension using the spectral geometry approach; the concluding Section {\bf 5} discusses further applications of our approach.

\section{Spectral dimension {\it a l\`{a}} Ho\v{r}ava}

To achieve our main goal - calculation of the spectral dimension using methods of spectral geometry, we will not need to specify the action (or dynamics) of the HL gravity (for a nice review of different HL models, see \cite{Blas:2010hb}). The dimension is completely determined by the geometry of a model. So we will use only the geometrical information: a foliated manifold with anisotropy between space and time. We start with a brief account on Ho\v{r}ava's approach to calculating the spectral dimension.\footnote{Through the text we are using different letters to denote dimension: $d_S,\ \mathrm{n},\ \mathrm{n}_a$ etc., to distinguish methods used for its calculation. At each point it should be clear what dimension is being discussed. In the end, all these calculations converge to produce the same answer.}

Partially motivated by the work \cite{Ambjorn:2005db}, where the spectral dimension of space-time was numerically calculated in the framework of casual dynamical triangulations (CDT), Ho\v{r}ava suggested \cite{Horava:2009if} that the same type of UV/IR behavior of the dimension should be observed in his model of gravity. The argument could be understood if one looks at the underlying structure of both models: in the CDT approach to quantum gravity one introduces a preferred causal structure while in HL-type models there exists a preferred foliation of space-time by three-dimensional geometries.

To actually calculate the spectral dimension, $d_S$, in \cite{Horava:2009if} a definition of $d_S$ was adopted that is particularly convenient for lattice models like CDT. Namely, let us consider a random walk on a lattice. Then the probability density $\rho({\bf w}, {\bf w'}; \tau)$ of walking from point ${\bf w}$ to point ${\bf w'}$ in time $\tau$ allows us to introduce the average return probability $P(\tau) = \rho({\bf w}, {\bf w'}; \tau)|_{{\bf w} = {\bf w'}}$. The spectral dimension of the lattice is defined by (cf. Eq.(\ref{dimW}) below)
\begin{eqnarray}
d_S = -2\frac{d\log P(\tau)}{d\log \tau} \ .\label{dims}
\end{eqnarray}

The same definition can be used for a manifold $\mathcal{M}$, but now instead of random walks one has to consider a diffusion process, i.e., in this case, $\rho({\bf w}, {\bf w'}; \tau)$ satisfies the diffusion equation:
\begin{eqnarray}
\frac{\partial}{\partial\tau}\rho({\bf w}, {\bf w'}; \tau) = \Delta_w \rho({\bf w}, {\bf w'}; \tau)\ \ ,\ \ \lim_{\tau\rightarrow +0}\rho({\bf w}, {\bf w'}; \tau)=\delta ({\bf w} - {\bf w'}) \ ,\label{difeq}
\end{eqnarray}
where $\Delta$ is a standard Laplacian on $\mathcal{M}$. Then, Eq.(\ref{dims}) produces the usual topological dimension of $\mathcal{M}$.

In the case of HL gravity, the manifold is supplemented with an additional structure - a foliation. In addition to this, the coordinates on leaves of the foliation (3d geometries) have scaling, different from the transversal coordinate (time):
\begin{eqnarray}\label{scaling}
\mathbf{x} \rightarrow \xi\mathbf{x}\ \ ,\ \ t \rightarrow \xi^z t \ .
\end{eqnarray}
This means that the right-hand side of Eq.(\ref{difeq}) is not a natural choice anymore: i) though the Laplacian in (\ref{difeq}) does respect the symmetry of a foliation, foliation preserving diffeos ($\mathrm{DIFF}_\mathcal{F}(\mathcal{M})$), it is too restrictive; ii) obviously, the usual Laplacian does not have a well-defined scaling under (\ref{scaling}). The idea is to keep (\ref{dims}) as the definition of dimension, but modify the RHS of (\ref{difeq}) in such a way that points i)-ii) are resolved. We will illustrate the choice of a generalized ``Laplacian'', $``\Delta"$, on an example of a flat space-time. Later, in Section {\bf 4}, we will argue (using a different approach) that the conclusion about the dimension of space-time remains true in a general case. It is clear, that the following choice resolves issues i)-ii) above
\begin{eqnarray}\label{ModLap}
``\Delta" := \frac{\partial^2}{\partial t^2} + (-1)^{z+1}\Delta^z \ ,
\end{eqnarray}
where $t$ is the (Euclidean) time, $\Delta$ is a $D$-dimensional spatial Laplacian and $z$ is the scaling exponent from Eq.(\ref{scaling}). The factor $(-1)^{z+1}$ is chosen so that $``\Delta"$ is elliptic in the generalized sense (see a discussion on ellipticity in Section {\bf 4}, Eq.(\ref{elliptic})). Using $``\Delta"$ instead of the standard Laplacian in the right-hand side of (\ref{difeq}), one easily obtains the average return probability $P(\tau)$ \cite{Horava:2009if}:
\begin{eqnarray}
P(\tau) = \rho({\bf w}, {\bf w'}; \tau)|_{{\bf w} = {\bf w'}} = \int \frac{d\omega d^D k}{(2\pi)^{D+1}}e^{\tau (\omega^2 + |\mathbf{k}|^{2z})} = \frac{const}{\tau^{(1+D/z)/2}} \ .\nonumber
\end{eqnarray}
Upon substituting this result into Eq.(\ref{dims}), we have for the spectral dimension
\begin{eqnarray}\label{dimH}
d_S = 1+\frac{D}{z}\ .
\end{eqnarray}
Then for $z=1$ (IR regime) $d_S = D+1$, i.e. usual dimension of the space-time, while for $z=3$ (UV regime) and $D=3$ we have $d_S = 2$.

To describe the transition from the IR regime to UV one, we have to use the complete Laplacian, which can be schematically written as
\begin{eqnarray}\label{completeL}
``\Delta" = \frac{\partial^2}{\partial t^2} + \Delta^3 + M_{2*}^2\Delta^2 + M_{1*}^4\Delta \ ,
\end{eqnarray}
where $M_{i*}$ are some scales. Then, it is clear that for the distances much greater than $1/M_*$ the space-time will look like the usual $4d$ one, while for the distances much shorter then $1/M_*$ it will ``flow'' to $2d$. More on this is given in the following and the concluding sections.

\section{Spectral dimension out of Dirac operators}

The task of the spectral geometry is extracting geometrical information about a manifold from the spectral properties of some differential operators. There are many references on spectral (or non-commutative) geometry \cite{GraciaBondia:2001tr},\cite{Connes:1994yd},\cite{Landi:1997sh},\cite{Madore:2000aq}. Here, we don't need the whole machinery, so we just briefly review the main idea. Roughly speaking, the spectral geometry is an elaborated answer to Mark Kac's question: ``Can one hear the shape of a drum?" \cite{Kac:1966xd}. Towards this end, one tries to obtain geometric information from the spectrum of some relevant operator (Laplacian or Dirac). For example, it is well known, (see, e.g., \cite{GraciaBondia:2001tr}) that the Riemannian geometry of a closed manifold, $\mathcal{M}$, can be completely recovered from the so-called spectral triple, ($\mathcal{A, H, D}$), where $\mathcal{A}=\mathcal{C}^\infty (\mathcal{M})$, $\mathcal{H}=L^2(\mathcal{M,S})$ - the Hilbert space of spinors on $\mathcal{M}$, and $\mathcal{D}=\D = i\gamma^\mu (\partial_\mu + \omega_\mu)$ - a standard Dirac operator on $\mathcal{M}$. Using more general spectral triples, ($\mathcal{A, H, D}$), (subject to some natural conditions) allows us to construct generalized geometries in purely algebraic way, including cases when the geometric construction either does not exist or is obscure. Moreover, because the Dirac operator is also very important from the physics point of view, this shows that choosing different Dirac operators, i.e. different physics, we change the {\it observable} geometry. By this we mean the following: the only way we can probe geometry is by performing some measurements and the outcome will depend on our choice of the Dirac operator.\footnote{Actually, if we define our physics through the spectral action principle (see the discussion in the conclusion), we can immediately see that the Dirac operator carries all the information not only about the geometry but about the physics as well.}

Here, we are particularly interested in how the information about the dimension of a manifold is obtained in this approach. The answer to this question is known by the name of Weyl's theorem:

{\it Let $\Delta$ be the Laplace operator on a closed Riemannian manifold $\mathcal{M}$ of dimension $\mathrm{n}$. Let $N_\Delta (\lambda)$ be the number of eigenvalues of $\Delta$, counting multiplicities, less then $\lambda$, i.e. $N_\Delta (\lambda)$ is the counting function
\begin{eqnarray}\label{counting}
N_\Delta (\lambda) := \#\{ \lambda_k (\Delta)\ :\ \lambda_k (\Delta)\leq \lambda\}\ .
\end{eqnarray}
Then
\begin{eqnarray}\label{Weyl}
\lim_{\lambda\rightarrow\infty}\frac{N_\Delta (\lambda)}{\lambda^{\frac{\mathrm{n}}{2}}}=\frac{Vol (\mathcal{M})}{(4\pi)^{\frac{\mathrm{n}}{2}}\Gamma(\frac{\mathrm{n}}{2}+1)}\ ,
\end{eqnarray}
where $Vol (\mathcal{M})$ is the total volume of the manifold $\mathcal{M}$.}

Weyl's theorem tells that from the spectrum of the Laplacian we can recover not only the dimension of the manifold, but also its volume. Here, we should comment why we switched from the Dirac operator (from the spectral triple) to the Laplace operator. In the case of the standard Dirac operator (which {\it must} be used to fully recover usual geometry) this is a consequence of the Lichnerowicz formula:
\begin{eqnarray}\label{Lichn}
\D^2 = \Delta^S + \frac{1}{4}R\ ,
\end{eqnarray}
where $R$ is the scalar curvature of $\mathcal{M}$ and $\Delta^S$ is the spinor Laplacian, which is closely related to the usual Laplacian. Then, the Dirac operator can be used instead of the Laplacian:
\begin{eqnarray}\label{WeylD}
\lim_{\lambda\rightarrow\infty}\frac{N_{|\D |} (\lambda)}{\lambda^{\mathrm{n}}}= \frac{2^m Vol (\mathcal{M})}{(4\pi)^{\frac{\mathrm{n}}{2}}\Gamma(\frac{\mathrm{n}}{2}+1)}\ ,
\end{eqnarray}
where $2^m$ is the dimension of spinors on $2m$- or $(2m+1)$-dimensional manifold.

In the case of generalized geometries, it is natural to use (\ref{Weyl}) or (\ref{WeylD}) as the definition of the dimension of the corresponding geometry. Using (\ref{Weyl}) (or (\ref{WeylD})) one can derive the formula for the dimension itself:
\begin{eqnarray}\label{dimW}
\mathrm{n} = 2\lim_{\lambda\rightarrow\infty}\frac{d \log N_{\Delta} (\lambda)}{d \log \lambda}=\lim_{\lambda\rightarrow\infty}\frac{d \log N_{|\D|} (\lambda)}{d \log \lambda}\ .
\end{eqnarray}
Notice strong resemblance between (\ref{dimW}) and the definition of the spectral dimension (\ref{dims}) used in \cite{Horava:2009if}. But now the definition of the dimension is more in the spirit of quantum physics: it is defined by the spectrum of a self-adjoint operator, $\Delta$ or $\D$. The strongest motivation to use (\ref{dimW}) as the definition of dimension comes from the Dixmier trace \cite{Dixmier}, which is a natural generalization of the integral over manifold. In the construction of this integral the role of the volume form is played by $|\D|^{-\mathrm{n}}$, where $\mathrm{n}$ is given by (\ref{dimW}) (see \cite{GraciaBondia:2001tr} for details and also some further comments in the next section).

After this brief introduction to the ideas of the spectral geometry, we are ready to calculate the spectral\footnote{Note that now the dimension is really {\it spectral}.} dimension of the HL space-time. We begin with a simple case of a flat geometry (as we did in Section {\bf 2} to demonstrate Ho\v{r}ava's approach) postponing the general consideration to the next section. For simplicity, let us consider the space with the topology of a torus, $T^{D+1} = S^1\times\cdots\times S^1$, where the radius of the first circle (corresponding to Euclidean time) is $T$ and the rest of $D$ circles have equal radii, $R$.\footnote{The choice of a compact space, e.g. $T^{D+1}$ in our case, is not too restrictive. From (\ref{Weyl})-(\ref{dimW}), we see that taking de-compactifying limit, $T,R\rightarrow\infty$, will not affect the dimension.} To recover the (flat) geometry of this manifold, including its dimension, one has to use the standard flat Dirac operator
\begin{eqnarray}
\D = i\gamma^\mu \partial_\mu \ . \nonumber
\end{eqnarray}
The Lichnerowicz formula (\ref{Lichn}) trivially produces
\begin{eqnarray}\label{D2L}
\D^2 = \Delta\mathds{1}_{2^m} \ ,
\end{eqnarray}
where $\Delta = -\delta^{ij}\partial_i \partial_j$ and $\mathds{1}_{2^m}$ is a unitary operator in the space of spinors. Instead of calculating the dimension of $T^{D+1}$ using (\ref{D2L}) we use the generalized Laplacian (\ref{ModLap}) (with the obvious redefinition $``\Delta" \rightarrow -``\Delta"$ to have a positive operator) obtaining the standard torus (IR regime) and HL torus (UV regime) as special cases. Then for the spectrum of $``\Delta"$, $sp(``\Delta")$, we have:
\begin{eqnarray}\label{spectrum}
sp(``\Delta") = \left\{ \frac{n_0^2}{T^2}+\frac{1}{R^{2z}}(\sum_{i=1}^{D}n_i^2)^z\ ,\ n_0,n_i\in\mathbb{Z}\right\} \ . \nonumber
\end{eqnarray}
Because we are interested only in the asymptotic behavior of the counting function (\ref{counting}), we can use integration instead of summation:
\begin{eqnarray}\label{counting1}
N_{``\Delta"}(\lambda)\sim \int_{-T\sqrt{\lambda}}^{T\sqrt{\lambda}}d n_0\int_0^{R\left(\lambda-\frac{n_0^2}{T^2}\right)^{1/2z}} \Omega_{D} \rho^{D-1}d \rho = \frac{\sqrt{\pi}\Omega_D(2\pi R)^D (2\pi T)}{D (2\pi)^{D+1}}\frac{\Gamma\left(\frac{D}{2z}+1\right)}{\Gamma\left(\frac{D}{2z}+\frac{3}{2}\right)}\lambda^{\frac{D}{2z}+\frac{1}{2}}\ ,
\end{eqnarray}
where $\Omega_D = Vol (S^{D-1})=\frac{2\pi^{D/2}}{\Gamma({D/2})}$.
\begin{itemize}
\item{$z=1$}\ :\ $N_{``\Delta"}(\lambda)\sim \frac{(2\pi R)^D (2\pi T)}{(2\pi)^\frac{D+1}{2}}\lambda^\frac{D+1}{2}$. Comparing this with (\ref{Weyl}), we immediately extract the standard volume of $T^{D+1}$, $Vol(T^{D+1})=(2\pi R)^D (2\pi T)$, and the classical dimension, $\mathrm{n}=D+1$.
\item{$z=3$, $D=3$}\ :\ $N_{``\Delta"}(\lambda)\sim \frac{(2\pi R)^3 (2\pi T)}{12(2\pi)^3}\lambda$. Using once again Weyl's theorem (\ref{Weyl}) (or (\ref{dimW})), we extract the spectral dimension of $T^4$ in the HL regime, $\mathrm{n}=2$.\footnote{To answer the question about volume, one has to look at the Dixmier trace with the `measure' $|\D|^{-\mathrm{n}}$. But our primary goal in this paper is the analysis of the spectral dimension. The discussion of the Dixmier trace in the context of the spectral action will be given elsewhere \cite{Pinzul}.}
\item{general $z$ and $D$}\ :\ $N_{``\Delta"}(\lambda)\sim const\times\lambda^\frac{1+D/z}{2}$, where $const$ depends on $T$ and $R$ but has no dependence on $\lambda$. Then, using Weyl's theorem one more time we see that the spectral dimension in a general case is $\mathrm{n}=1+\frac{D}{z}$, in perfect agreement with Ho\v{r}ava's result (\ref{dimH}).
\end{itemize}

\begin{figure}[htb]
\begin{center}
\leavevmode
\includegraphics[scale=0.7]{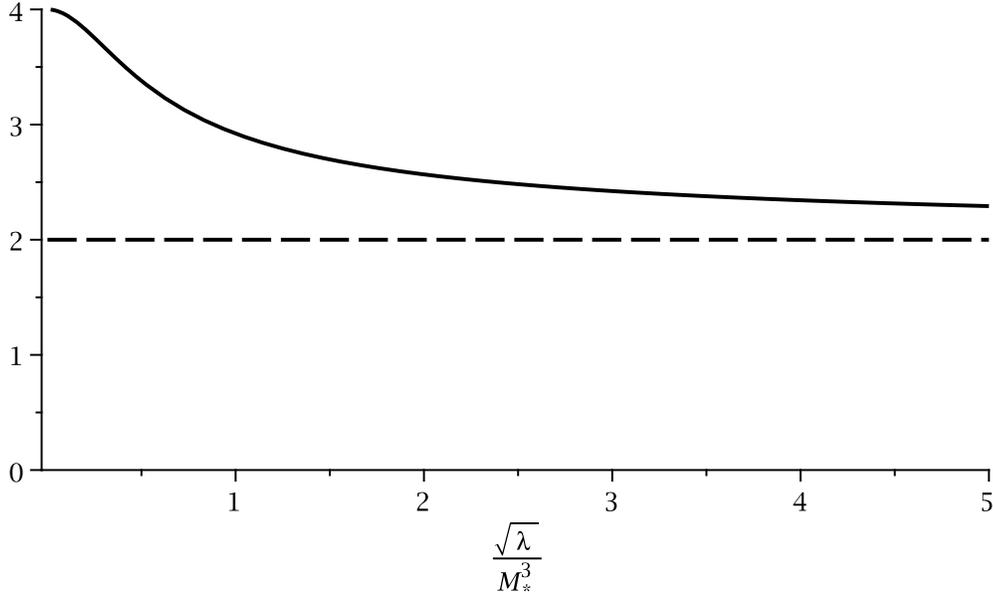}
\end{center}
\caption{The example of the transition from the IR to UV regime for $D=3$ and $z=3$. The spectral dimension is plotted as a function of the cut-off for the specific choice of the constants in (\ref{completeL}): $M^2_{2*}/2=M^2_{1*}=M^2_*$.}
\label{fig:scaling}
\end{figure}

What about the transition between the IR regime, which corresponds to $z=1$, and the UV regime with $z=3$? Of course, to answer this question we have to know the exact form of the complete generalized Laplacian. But some conclusions can be drawn without using this knowledge. Here, we will continue to use the case of a flat space-time and will comment on a general case in the next section. We want to analyze the spectral dimension of the generalized Laplacian (\ref{completeL}) (though from our consideration it will be obvious that only the form of the second, UV, and last, IR, terms is important). So we want to introduce a spectral dimension depending on energy (through the spectrum cut-off $\lambda$), $\mathrm{n}_\lambda$, which is defined by exactly the same formula (\ref{dimW}) but now with the complete Laplacian (\ref{completeL}). The spectrum of $``\Delta"$ now takes the form
\begin{eqnarray}\label{spectrum1}
sp(``\Delta") = \left\{ \frac{n_0^2}{T^2}+\sum_{k=1}^z\frac{M_{k*}^{2(z-k)}}{R^{2k}}(\sum_{i=1}^{D}n_i^2)^k\ ,\ n_0,n_i\in\mathbb{Z}\right\} \ .
\end{eqnarray}
Using (\ref{dimW}) and (\ref{spectrum1}), trivial algebra leads to the following result:
\begin{eqnarray}\label{spectraldimgen}
\mathrm{n}_\lambda = 1+ \frac{D}{z} + 2\lambda \frac{d}{d\lambda} \ln \idotsint\limits_V \rho^{D-1}d\rho dn \ ,
\end{eqnarray}
where $V=\{{n^2+\sum\limits_{k=1}^z\tilde{M}_k^{2(z-k)}\rho^{2k}\leq 1}\}$ and $\tilde{M}^2_k = \frac{M^2_{k*}}{\lambda^{1/z}}$. Performing integration over $n$, we obtain
\begin{eqnarray}\label{spectraldimgen1}
\mathrm{n}_\lambda = 1+ \frac{D}{z} + \frac{1}{z} \int\limits_0^{\rho_0} \frac{{\sum\limits_{k=1}^z (z-k)\tilde{M}_k^{2(z-k)}\rho^{2k}}}{\sqrt{{1-\sum\limits_{k=1}^z\tilde{M}_k^{2(z-k)}\rho^{2k}}}} \rho^{D-1}d\rho \left/ \int\limits_0^{\rho_0} {\sqrt{{1-\sum\limits_{k=1}^z\tilde{M}_k^{2(z-k)}\rho^{2k}}}}\ \rho^{D-1}d\rho\ , \right.
\end{eqnarray}
where $\rho_0$ is the solution of the equation $\sum\limits_{k=1}^z\tilde{M}_k^{2(z-k)}\rho^{2k}=1$. This formula is easily analyzed in two important limits:
\begin{itemize}
\item{IR}\ : The exact meaning of IR is that $\lambda\ll M^{2z}_{i*}$. In this limit $\rho_0\approx 1/\tilde{M}_1^{z-k}$ and all integrals are trivially calculated leading to the effective $z=1$ behavior, cf. the first case after (\ref{counting1}),
    $$
    \mathrm{n}_\lambda \approx 1+D \ .
    $$
\item{UV}\ : In this case $\lambda\gg M^{2z}_{i*}$. This corresponds to $\rho\approx 1$. In this case, it is easy to see that the integral term in (\ref{spectraldimgen1}) is of the order of $\mathcal{O}(\frac{M^2_{i*}}{\lambda^{1/z}})$. Then, we reproduce the UV spectral dimension, which corresponds to the general case after (\ref{counting1})
    $$
    \mathrm{n}_\lambda \approx 1+\frac{D}{z} \ .
    $$
\end{itemize}
From this discussion it should be clear that both, UV and IR, limits do not depend on the exact form of the complete Laplacian. What really depends on it are the details of the transition. To give a flavor of how this transition might look like, in Fig.(\ref{fig:scaling}) we plot the dependence of the effective spectral dimension on the cut-off scale $\lambda$ for some choice of $M_{i*}$. It is interesting to see that the departure from the classical dimension is rather rapid. But because typically we would think of $M_*$ as the Plank scale, of course, this is not something that we could observe experimentally.

In this section, we have seen that the spectral geometry approach based on Weyl's theorem produces exactly the same dimension (at least in the case of a flat space-time) as the dimension based on random walks (or diffusion process). This is not a big surprise, because the same operator, $``\Delta"$, plays a crucial role in both approaches. But, as we discuss in Section {\bf 5}, the spectral geometry methods, from our point of view, seem more physically motivated, especially in view of possible further applications. Let us first generalize our discussion to more general, non-flat geometries.

\section{Meromorphic continuation of a $\zeta$-function}

Let us define a generalized $\zeta$-function for a Laplace operator on some closed manifold $\mathcal{M}$
\begin{eqnarray}\label{zeta}
\zeta_\Delta (s):= \Tr (\Delta^{-s})\ .
\end{eqnarray}
Then using Weyl's theorem one can easily see that (\ref{zeta}) is well defined for $Re (z)> \frac{\mathrm{n}}{2}$, where $\mathrm{n}$ is, as usual, the dimension of the manifold $\mathcal{M}$. In \cite{Higson}, it is proven that (\ref{zeta}) has a meromorphic extension to the complex plane $\mathbb{C}$ with the only poles at the points\footnote{Actually, in \cite{Higson} the more general result is proven: if $A$ is a linear differential operator on $\mathcal{M}$ of order $q$, then $\Tr (A\Delta^{-s})$ extends to a meromorphic function on the complex plane $\mathbb{C}$ with the only poles at the points
$$
\frac{\mathrm{q+n}}{2}\ ,\ \frac{\mathrm{q+n-1}}{2}\ ,\ \cdots\ \frac{\mathrm{q+n-k}}{2}\ ,\ \cdots
$$
For our purpose, the $q=0$ case will be sufficient.}
\begin{eqnarray}\label{poles}
\frac{\mathrm{n}}{2}\ ,\ \frac{\mathrm{n-1}}{2}\ ,\ \cdots\ \frac{\mathrm{n-k}}{2}\ ,\ \cdots
\end{eqnarray}

From (\ref{poles}), we see that the information about the dimension of a manifold is fully recovered from the pole structure of $\zeta$-function (\ref{zeta}). What is the connection between this method and the one we used in the previous section? We have already mentioned that the correct way to think about dimension (adopted in non-commutative geometry \cite{Connes:1994yd}) is the following: dimension is a number $\mathrm{n}\in \mathbb{R}$ such that the Dixmier trace\footnote{The Dixmeir trace, $\Tr^+$, is defined for a special class of operators such that the usual trace diverges logarithmically.} $\Tr^+ |\mathcal{D}|^{-\mathrm{n}}$, where $\mathcal{D}$ is a Dirac operator, exists. Now, using Connes' trace theorem \cite{Connes:1988ym} (or rather some part of it), we have
 \begin{eqnarray}
\Tr^+ |\mathcal{D}|^{-\mathrm{n}} = \Tr^+ \Delta^{-\frac{\mathrm{n}}{2}} = Res |_{s=\frac{\mathrm{n}}{2}} \Tr \Delta^{-s} \ , \nonumber
\end{eqnarray}
which is exactly the contribution of the first pole in (\ref{poles}). Then we can take this as a \textit{definition} in general: the first pole of $\zeta_\Delta (s)= \Tr (\Delta^{-s})$ is equal half of the dimension. We will apply this definition to generalized geometries (i.e., defined by generalized Laplacians, e.g., by $``\Delta"$). Here, we will not need all these technicalities and gave the extensive comment above just to clarify the connection between two types of calculations. All needed details can be found in \cite{GraciaBondia:2001tr},\cite{Connes:1994yd}.

Motivated by the work of Connes and Moscovici \cite{Connes}, $\zeta$-functions for more general then the standard Laplacian operators were analyzed in \cite{Higson}. In particular, the construction for a foliation of the co-dimension $p$ was considered. Here, we give just enough details to construct the generalization to a foliation of the HL type, for the complete treatment, see \cite{Higson}.

Let us consider an integrable smooth sub-bundle $\mathcal{F}$ of the tangent bundle $T\mathcal{M}$. Then, it defines a foliation of $\mathcal{M}$ of some co-dimension $p$. The algebra of linear differential operators on $\mathcal{M}$ has a natural filtration defined by the order of differential operators. Now, in the presence of the additional structure (foliation), we can define another filtration assigning non-standard order to operators:
\begin{itemize}
\item[1)]\ $order(f)=0\ \ \forall f\in C^\infty(\mathcal{M})$
\item[2)]\ $order(X)\leq 1\ \ \forall X$ - $C^\infty$ vector field on $\mathcal{M}$ tangent to $\mathcal{F}$
\item[3)]\ $order(X)\leq 2\ \ \forall X$ - $C^\infty$ vector field on $\mathcal{M}$ (not necessarily tangent to $\mathcal{F}$)
\end{itemize}

Using the more physical language, we see that the filtering based on 1)-3) in the case of $p=1$ (the transversal coordinate being time) is nothing but anisotropic scaling (\ref{scaling}) with $z=2$: $\mathbf{x} \rightarrow \xi\mathbf{x},\ t \rightarrow \xi^2 t$. After this comment, the generalization to the HL case ($z=3$) is almost trivial: instead of the order prescription 3) we will introduce the more general one:
\begin{itemize}
\item[3\'{})]\ $order(X)\leq z\ \ \forall X$ - $C^\infty$ vector field on $\mathcal{M}$ (not necessarily tangent to $\mathcal{F}$)
\end{itemize}
and proceed with this general case. Any differential operator $X$ of the generalized order less or equal $k$, $order(X)\leq k$, can be written in the following form
\begin{eqnarray}\label{X}
X=\sum_{\|\alpha\|=k}X_{\alpha_1\cdots\alpha_\mathrm{n}}(x)\frac{\partial^{\alpha_1\cdots\alpha_\mathrm{n}}}{\partial x^{\alpha_1\cdots\alpha_\mathrm{n}}}+\sum_{\|\alpha\|<k}\tilde{X}_{\alpha_1\cdots\alpha_\mathrm{n}}(x)\frac{\partial^{\alpha_1\cdots\alpha_\mathrm{n}}}{\partial x^{\alpha_1\cdots\alpha_\mathrm{n}}}\ ,
\end{eqnarray}
where $\|\alpha\|:= \sum_{i=1}^{\mathrm{n}-p}\alpha_i + z\sum_{i=\mathrm{n}-p+1}^{\mathrm{n}}\alpha_i$. Using the decomposition (\ref{X}) of $X$, we can introduce notion of the generalized ellipticity. We call $X$ to be an elliptic operator if
\begin{eqnarray}\label{elliptic}
\left|\sum_{\|\alpha\|=k}X_{\alpha_1\cdots\alpha_\mathrm{n}}(x)\xi^{\alpha_1\cdots\alpha_\mathrm{n}}\right|\geq \epsilon (x)\left(\sum_{i=1}^{\mathrm{n}-p}|\xi_i|^2 + \sum_{i=\mathrm{n}-p+1}^{\mathrm{n}}|\xi_i|^{2z}\right)
\end{eqnarray}
for some $\epsilon (x)>0$ and all $\xi\in\mathbb{R}^\mathrm{n}$. When $z=1$ this reduces to the usual definition of an elliptic operator and when $z=2$ we are back to \cite{Higson}. Using this definition of ellipticity, we see that the generalized Laplacian (\ref{ModLap}), which we used to calculate the spectral dimension, is really elliptic and of the order $order (``\Delta")=2z$. It is now clear, that any differential operator that respects the foliation preserving symmetry, $\mathrm{DIFF}_\mathcal{F}(\mathcal{M})$, will be of the form (\ref{X}). Then the result we are going to obtain will be valid for a general, non-flat, foliated manifold, e.g. HL space-time.

We want to use the relation between the first pole of the $\zeta$-function (\ref{zeta}) to the dimension of $\mathcal{M}$ as the definition of dimension in the case of generalized geometries, i.e. defined by the generalized Laplacians. This is so called {\it analytic} dimension, which, as we have already seen in the case of the usual manifold, coincides with the spectral dimension (\ref{dimW}). We start with defining a $\zeta$-function for the generalized Laplacian, which is now an operator of order $2z$
\begin{eqnarray}\label{zeta1}
\zeta_{``\Delta"} (s):= \Tr (``\Delta"^{(-s)})\ .
\end{eqnarray}
Using the analysis that parallels the one in \cite{Higson}, we arrive at the result that $\zeta_{``\Delta"} (s)$ can be extended to a meromorphic function on the complex plane with the only poles at
\begin{eqnarray}\label{poles1}
\frac{\mathrm{n}-p+zp}{2z}\ ,\ \frac{\mathrm{n}-p+zp-1}{2z}\ ,\ \cdots\ \frac{\mathrm{n}-p+zp-k}{2z}\ ,\ \cdots
\end{eqnarray}
Requiring that (\ref{poles1}) has the form (\ref{poles}), where now $\mathrm{n}$ should be taken as $\mathrm{n}_a$ - analytic dimension, we can calculate the (analytic) dimension of the generalized geometry:
\begin{eqnarray}\label{dimA}
\frac{\mathrm{n}-p+zp}{2z}=\frac{\mathrm{n}_a}{2}\ \Rightarrow \ \mathrm{n}_a = \frac{\mathrm{n}-p+zp}{z}\ . \nonumber
\end{eqnarray}
Specifying to the case of $p=1$ and writing $\mathrm{n}=D+1$ we obtain
\begin{eqnarray}\label{dimA1}
\mathrm{n}_a = 1 + \frac{D}{z}\ .
\end{eqnarray}
This is exactly the dimension (\ref{dimH}) in Ho\v{r}ava's approach as well as the dimension that we obtained for the flat torus using Weyl's theorem. This finishes the proof that the dimension of an arbitrary (closed) foliated manifold, where the filtration is given by a non-standard order prescription, specified by the anisotropic scaling is given by (\ref{dimA1}).

At the end of this section we would like to discuss, as we did for the case of a flat space-time, the IR/UV transition. The way the analytical dimension was defined - through the poles of the generalized $\zeta$-function (\ref{zeta1}), it is the dimension in UV. This is due to the fact that the trace in (\ref{zeta1}) is over the whole Hilbert space, i.e. it involves the {\it whole} spectrum of the generalized Laplacian $``\Delta"$. The situation is in complete analogy with the consideration in Section {\bf 3}: if we would have used the definition of the spectral dimension as in (\ref{dimW}), i.e. in the limit, we would have obtained exactly the UV dimension. Instead, we introduced some cut-off $\lambda$ and were able to analyze the scale dependent dimension (see the next section on the physical meaning of this scale dependence). It is pretty obvious that the same can be done in a general case of a generalized Laplacian on a curved space-time. In other words, instead of (\ref{zeta1}) we should consider a "truncated $\zeta$-function"
\begin{eqnarray}\label{zetatrunc}
\zeta^\lambda_{``\Delta"} (s):= \Tr_\lambda (``\Delta"^{(-s)})\ ,
\end{eqnarray}
where $\Tr_\lambda$ is the trace over the subspace of the full Hilbert space, corresponding to eigenvalues of $``\Delta"$ less than or equal to $\lambda$. So, for example, $\zeta^\lambda_{``\Delta"} (0)$ is just the counting function $N_{``\Delta"}(\lambda)$ from (\ref{counting1}). Obviously, $\zeta^\lambda_{``\Delta"} (s)$, being a finite sum for any finite $\lambda$ does not have any singularities - poles exist only in the exact limit $\lambda \rightarrow \infty$. From this definition, it should be clear that, as in the flat case, the dimension in UV does not depend on the details of the generalized Laplacian and completely determined by the anomalous scaling $z$ as in (\ref{dimA1}), i.e. the first, UV, term in (\ref{X}). On the other hand, when $\lambda\ll M^{2z}_*$ for some characteristic scale $M_*$, i.e. in IR, we will see the development of the wouldbe singularity at $s=\frac{1+D}{2}$. But increasing the cut-off, the position of this apparent singularity will move from $s=\frac{1+D}{2}$ to $s=\frac{z+D}{2z}$, where it will become a pole when $\lambda \rightarrow \infty$. Of course, the details of this "flow", as in the case of the flat space-time, will depend on the actual form of the generalized Laplacian.

\section{Discussion and conclusion}

The consideration above could appear too formal. But let us look at this from a physical point of view. A Dirac operator and the generalized Laplacian related to it are self-adjoint operators on a Hilbert space of spinors. Quantum Mechanics teaches us that they are observables (related to energy) and their spectrum is what we, in principle, would measure by setting up an appropriate experiment. We can use the spectrum to calculate the dimension of a space-time using the generalization of Weyl's theorem (\ref{Weyl}). Note that at this point Weyl's theorem stops being just a mathematical fact but turns into the functional way of measuring dimension: experimental data plus Weyl's theorem equals space-time dimension. But any real experiment will be able to measure eigenvalues only up to some maximal value determined by the resolution of our experiment. Now, if this value is well below some scale $M_*$ defined in (\ref{completeL}) the spectrum will look like a spectrum of just the last term in (\ref{completeL}). Then we will have to conclude that our space-time corresponds to the first case after Eq.(\ref{counting1}), i.e. $z=1$, and as a consequence is four dimensional. But if, on the other hand, we can measure the spectrum beyond this scale $M_*$,\footnote{Actually, as we can see from Fig.(\ref{fig:scaling}), the deviation from the classical dimension $d=4$ starts well below the $M_*$ scale (though still well above any energy we can probe if $M_*$ is of the order of the Planck scale).} we will start noticing that the counting function (\ref{counting1}) starts to deviate from $\lambda^2$ behavior and behaves more and more like $\lambda$. At this point, we are forced to conclude that the actual (microscopic) dimension of our space-time is two. Here, as we briefly discussed in Section \textbf{2}, it is very important to distinguish between the topological dimension of the space-time, which still remains to be 4, and the spectral, i.e. physical, dimension defined by the Dirac operator. It is the latter that we call the microscopic dimension because it is the dimension seen by the physics at the microscopic, i.e. much greater then $M_*$, scale.

Is it possible to tell more about the transition from $4d$ to $2d$ in addition to our previous discussion? Let us again consider the generalized Laplacian (\ref{ModLap}). Using the foliation structure of the space-time, we can ask: what is the dimension of leaves or space? To answer this question, let us use Weyl's theorem (\ref{Weyl}) but only for the space part of (\ref{ModLap}), $\Delta_{space}=\Delta^z$. Then we immediately see that the corresponding counting function (\ref{counting}) behaves as
 \begin{eqnarray}\label{countspace}
N_{\Delta^z}(\lambda) \sim \lambda ^{\frac{D}{2z}}\ .
\end{eqnarray}
From (\ref{countspace}) we deduce the spectral dimension of space, $d_{space}=\frac{D}{z}$. Then we have the following interpretation of (\ref{dimH}) or (\ref{dimA1}): The dimension of time always remains $d_{time}=1$ (this corresponds to ``$1$'' in (\ref{dimH}), (\ref{dimA1})), while the dimension of space ``flows'' from $d_{space}=3$ in IR to $d_{space}=1$ in UV ($\frac{D}{z}$ term in (\ref{dimH}), (\ref{dimA1})).

Now we turn our attention to the advantages, which our approach can provide. As we mentioned in the introduction, one of the serious problems of HL models, from our point of view, is uncontrollable growth of the number of terms in the action: one has to allow more and more terms to deal with the phenomenology. On the other hand, there is a very powerful construction, called \textit{spectral action} \cite{Chamseddine:1996zu}, which provides a recipe on how to construct an action for a physical theory out of its Dirac operator. The geometric part of this action is just a trace of the Dirac operator:
\begin{eqnarray}\label{spectralA}
S_{geom}\sim \Tr f\left(\frac{\mathcal{D}}{\Lambda}\right)\ ,
\end{eqnarray}
where $\Lambda$ is some scale, which may or may not be related to $M_*$ and $f$ is some cut-off function. It is known, that (\ref{spectralA}) (plus the term we discuss below, Eq.(\ref{fermions})) fully reproduces Standard Model coupled to General relativity \cite{Chamseddine:2008zj}. Why is this an advantage in the case of HL gravities? The point is that there is much less freedom in choosing Dirac operator then the action itself: after satisfying all the requirements to be compatible with the foliation structure, anisotropic scaling and renormalizability, the Dirac operator will be fixed up to several dimensionful constants (like $M_*$). Then (\ref{spectralA}) will produce an action where we will have much better control over coupling constants (at least on the classical level).

It is very important, that having a Dirac operator, $\mathcal{D}$, and corresponding Hilbert space of fermions, $|\psi\rangle\in\mathcal{H}$, we immediately can construct the correct coupling of fermions to HL gravity:
 \begin{eqnarray}\label{fermions}
S_{ferm}\sim \langle\psi|\mathcal{D}|\psi\rangle\ .
\end{eqnarray}

From this consideration, we can see that the construction of the Dirac operator and the Hilbert space (on which the Dirac operator is represented) is the next logical step in our approach. This is the subject of our current research \cite{Pinzul}.

\section*{Acknowledgement}
The author acknowledges partial support of CNPq under grant no.308911/2009-1.

\end{document}